\documentclass[prd, showpacs, twocolumn,superscriptaddress]{revtex4}

\usepackage{graphicx}
\usepackage[normalem]{ulem}	
\usepackage[usenames]{color}    

\def\lsim{\raise0.3ex\hbox{$\;<$\kern-0.75em\raise-1.1ex
\hbox{$\sim\;$}}}
\def\gsim{\raise0.3ex\hbox{$\;>$\kern-0.75em\raise-1.1ex
\hbox{$\sim\;$}}}

\let\vev\VEV

\begin{document}
\title{Revisiting the Triangulation Method for Pointing to 
Supernova and Failed Supernova with Neutrinos}
\author{T.~M\"uhlbeier}
\email{muhlbeier@fis.puc-rio.br} 
\affiliation{Departamento de F\'{\i}sica,
  Pontif{\'\i}cia Universidade Cat{\'o}lica do Rio de Janeiro,
  C. P. 38071, 22452-970, Rio de Janeiro, Brazil}

\author{H.~Nunokawa}
\email{nunokawa@puc-rio.br} 
\affiliation{Departamento de F\'{\i}sica,
  Pontif{\'\i}cia Universidade Cat{\'o}lica do Rio de Janeiro,
  C. P. 38071, 22452-970, Rio de Janeiro, Brazil}
\author{R.~Zukanovich Funchal} 
\email{zukanov@if.usp.br}
\affiliation{Instituto de F\'{\i}sica, Universidade de S\~ao Paulo,
  C.\ P.\ 66.318, 05315-970 S\~ao Paulo, Brazil} 
\date{Augst 13, 2013}


\begin{abstract} 
In view of the advent of large-scale neutrino detectors such as
IceCube, the future Hyper-Kamiokande and the ones proposed for the
Laguna project in Europe, we re-examine the determination of the
directional position of a Galactic supernova by means of its neutrinos
using the triangulation method.
We study the dependence of the pointing accuracy on the arrival time 
resolution of supernova neutrinos at different 
detector locations.
For a failed supernova, we expect better results due to 
the abrupt termination of the neutrino emission which allows one to 
measure the arrival time with higher precision. 
We found that for the time resolution of $\pm$ 2 (4) ms, the supernova can be
located with a precision of $\sim$ 5 (10)$^\circ$ on the declination and
of $\sim$ 8 (15)$^\circ$ on the right ascension angle, if we combine the
observations from detectors at four different sites.
\end{abstract} 

\maketitle

\section{Introduction}
\label{sec:intro}

The observation of neutrinos coming from the next Galactic supernova
(SN) driven by gravitational core collapse (hereafter, SN implies the
one caused by the gravitational collapse) is expected to provide very
interesting information on the dynamics of the process, namely, how
these stars explode and form black holes, see for instance
Ref.~\cite{Bethe:1990mw,Janka:2012wk}.  Moreover, it may also shed
light on some unknown neutrino properties such as the neutrino mass
ordering, see for e.g. ~\cite{Dighe:1999bi,Serpico:2011ir}.

Since neutrinos can break free from the dense region of the star from
which photons cannot escape, 
they will be the first messengers from the sky to inform us
the occurrence of the gravitational collapse.  
Indeed, it might be possible that the next Galactic SN cannot be
located by optical observations due to obscuration. If so, observing
neutrinos may be the only way to access its direction in the sky,
apart from the possible simultaneous detection of gravitational
waves~\cite{Pagliaroli:2009qy}.

\begin{figure*}[!t]
\begin{center}
\hglue -1.0cm
\includegraphics[width=0.78\textwidth]{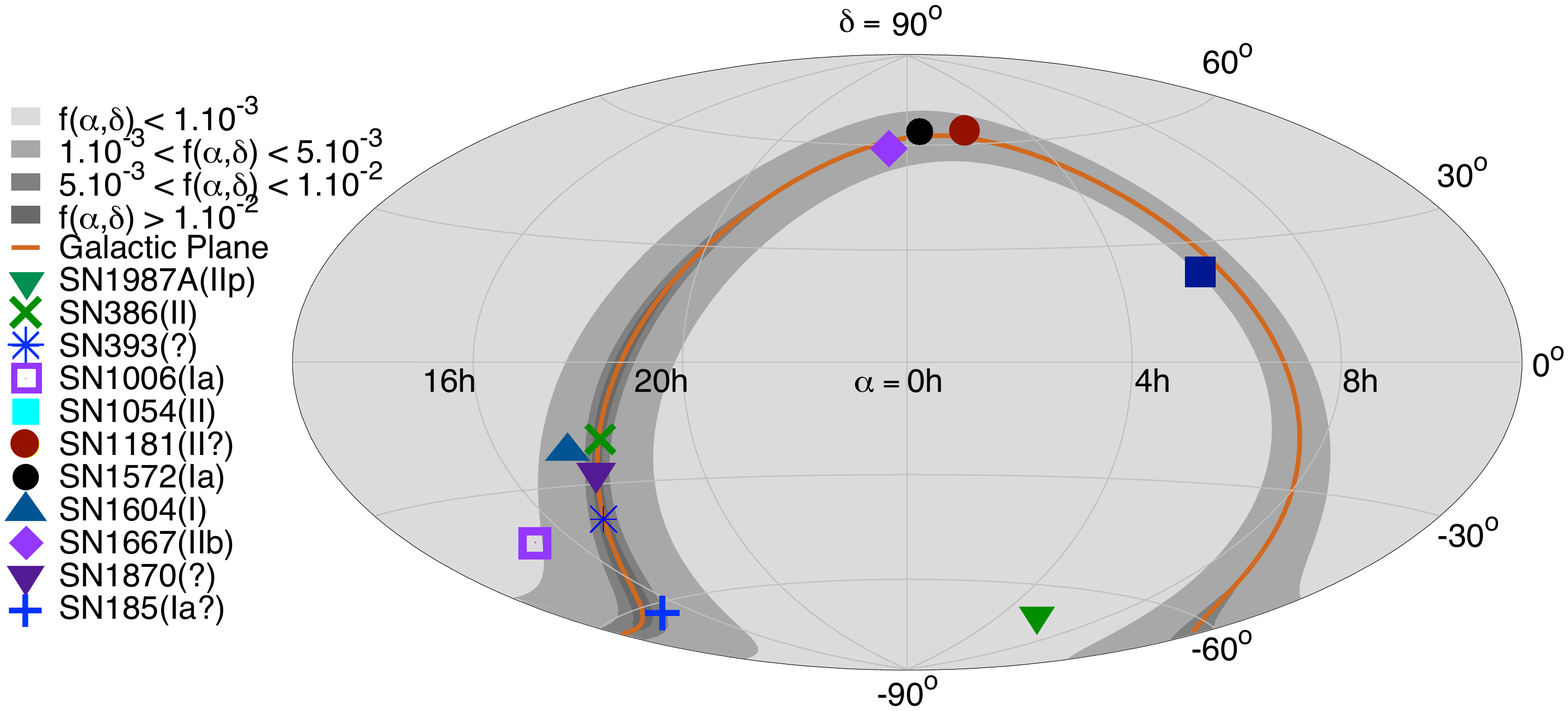}
\end{center}                          
\vglue -0.9cm
\caption{Expected SN probability distribution $f(\alpha,\delta)$
based on the model considered in Ref.~\cite{Mirizzi:2006xx}, 
shown in the plane of 
the equatorial coordinates $\alpha-\delta$ using the Hammer projection.
Different contrast of the colors reflects the difference 
in probabilities as indicated in the legend.
The position of the Galactic plane in the sky is indicated by 
the red curve.  The location of the historical Galactic SN explosions  
are also shown with their type, when known, in parentheses. }
\label{fig:SN-dist}
\end{figure*}

The possibility of determining the direction of a Galactic SN by
merely using its neutrinos, has been investigated in the
past~\cite{Burrows:1991kf,Habig-etal-neutrino1998,Beacom:1998fj,Apollonio:1999jg,Ando:2001zi,Tomas:2003xn,Scholberg:2009jr}.
Most of the authors considered neutrino electron elastic scattering events 
in a water Cherenkov detector in order to determine the SN 
direction~\cite{Burrows:1991kf,Beacom:1998fj,Ando:2001zi,Tomas:2003xn}.
According to Ref.~\cite{Tomas:2003xn}, for a SN at 10 kpc, 
the pointing accuracy is $\sim 8^\circ$ at 95\% C.L. 
if the Super-Kamiokande detector is considered. 
This can be further improved to $\sim 3^\circ$ if gadolinium is added to  
water~\cite{Beacom:2003nk}, allowing to tag neutrinos from the 
inverse beta decay background. A megaton size water Cherenkov 
detector using this technique may be able to increase the pointing 
precision to $\sim 1^\circ$~\cite{Tomas:2003xn}. 

On the other hand, the method of the arrival-time triangulation,
previously discussed in
Refs.~\cite{Burrows:1991kf,Habig-etal-neutrino1998,Beacom:1998fj}, was
readily dismissed due to the low precision on the arrival time of SN
neutrinos expected mainly because the available detectors at that time
were too small to register enough statistics for such a purpose.

We are now, however, entering a new era of large-scale detectors with
IceCube currently working in the South Pole~\cite{icecube}, the
proposals of Hyper-Kamiokande in Japan~\cite{HK} and of the European
detectors which will be built in the Pyh\"asalmi mine in
Finland~\cite{laguna}.
In view of this new trend, it is timely to revisit the usefulness of
neutrino triangulation using big detectors in different continents, as
suggested in Ref.~\cite{Halzen:2009sm}.

According to ~\cite{Halzen:2009sm}, the IceCube detector can determine the 
arrival time of SN neutrinos with an uncertainty of 
$\pm$ 3.5 ms at 95 \% C.L.. 
In the case of a so-called failed SN,  where a black hole is formed 
while the neutrino flux is still measurably high~\cite{Sumiyoshi:2006id}, 
one expects the neutrino signal to terminate abruptly.
As this sharp transition is expected to take
place in $\lsim $ 0.5 ms~\cite{Beacom:2000qy}
the end-point of the neutrino spectrum can also be used for triangulation. 

Since the observation of the arrival of SN neutrinos by various
detectors will be a valuable tool to alert astronomers about the
occurrence of the star collapse ~\cite{Antonioli:2004zb} allowing them
to observe the light curve as early as possible or the formation of a
black hole (in the case of a failed SN), it is important to explore
different approaches to reconstruct the location of 
the SN as well as the failed SN in the sky.

\section{Triangulation Method}
\label{sec:triangulation}

The distribution of SN in the Milky Way is expected to be
concentrated in the Galactic disc. 
For the sake of discussion, let us consider 
the same SN distribution considered 
in Refs.~\cite{Mirizzi:2006xx,Machado:2012ee}.
In Fig. \ref{fig:SN-dist} we show the expected SN distribution
$f(\alpha,\delta)$, in the plane of equatorial coordinates
$\alpha-\delta$ where $\alpha$ and $\delta$ are, respectively, right
ascension and declination, and $f(\alpha,\delta) \, d\alpha \, \cos
\delta \, d\delta $ corresponds to the probability to find a SN in the
sky in the interval between ($\alpha, \alpha+d \alpha$) and $(\delta,
\delta +d \delta)$.  The distribution function $f(\alpha,\delta)$ is
normalized, as in \cite{Mirizzi:2006xx}, such that $\int d \alpha \int
\, \cos \delta \, d\delta \, f(\alpha,\delta) = 1$ with $\alpha$ and
$\delta$ given in radian.
In this figure, we also show the location of the historical 
Galactic SN and SN1987A explosions.

\begin{figure*}[!t]
\begin{center}
\hglue -1.6cm
\includegraphics[width=0.84\textwidth]{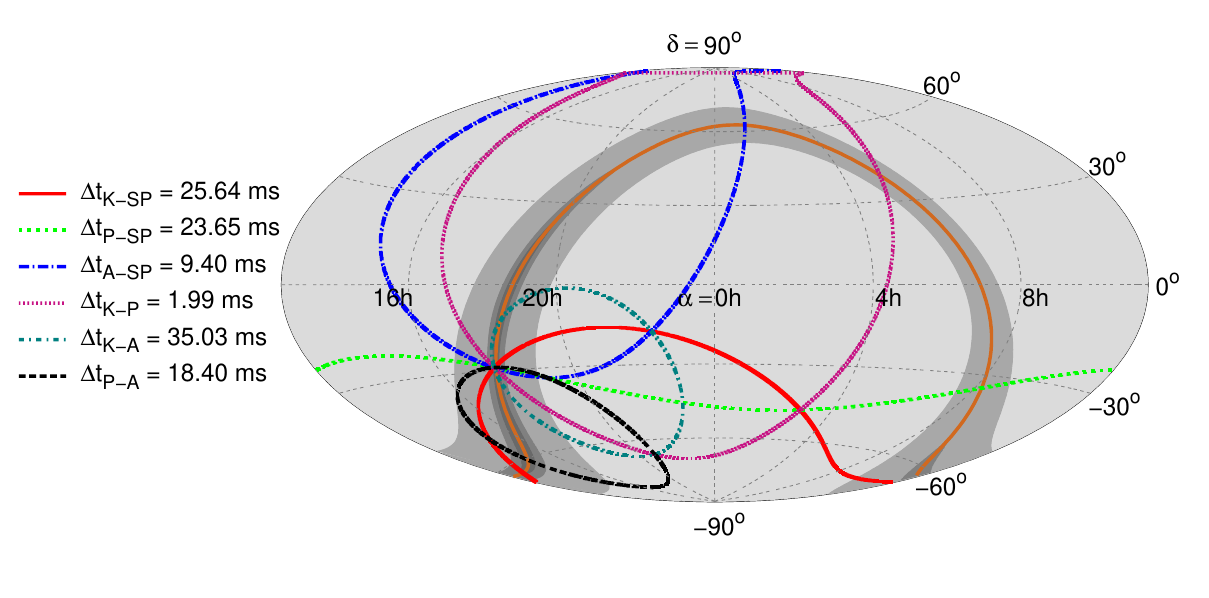}
\end{center}                          
\vglue -0.9cm
\caption{
Possible solutions for the SN direction ($\alpha$,
 $\delta$) consistent with a certain difference of the arrival 
time determined  by the combinations of detectors located 
at two different sites. Here the true (input) position of the SN is 
assumed to be the Galactic center, with $\alpha$ = 17h42m27s and 
$\delta$ = -28$^\circ$55'.  It is assumed that SN neutrinos are
detected at the Earth on the vernal point on March 20$^{th}$, 2000 at 12:00
UTC. 
We consider  four sites: Kamioka, Pyh\"asalmi, ANDES 
and the South Pole, indicated by the labels K, P, A and SP, 
respectively.}
\label{fig:two-detectors}
\end{figure*}

Let us consider two arbitrary detector sites {\boldmath $x_i$}
and {\boldmath $x_j$} on the Earth and define 
the displacement vector as {\boldmath $d_{ij} \equiv x_i-x_j$}, 
and denote the SN direction in the sky 
by the unit vector {\boldmath $n$}. 
Then the difference of the arrival time of SN neutrino signals 
between two detectors, $\Delta t_{ij}\equiv t_i - t_j$, 
is given by 
\begin{equation}
\Delta t_{ij} = \mbox{\boldmath $d_{ij}$}\cdot \mbox{\boldmath $n$}/c,
\label{eq:time-difference} 
\end{equation}
where $c$ is the speed of light in vacuum. 
Here we will ignore the possible time delay due to the 
neutrino mass which can be estimated as
\begin{equation}
\Delta t_{\text{mass}} \simeq 0.6  
\left[\frac{D}{10 \ \text{kpc}} \right] 
\left[\frac{m_\nu}{0.1\ \text{eV}} \ \frac{10\ \text{MeV}}{E}\right]^2 
\ \text{ms},
\end{equation}
where $D$ is the distance to the SN and $m_\nu$ is the neutrino mass. 

In this work, for the purpose of illustration of this method, 
we consider up to four different detector sites on Earth, 
namely, Kamioka, the South Pole,  Pyh\"asalmi and 
ANDES (Agua Negra Deep Experiment Site) ~\cite{Bertou:2012fk,Bertou:2013oda} 
(see also \cite{ANDES-WEB}). 
This is because four is the minimum number of detector positions
needed to uniquely determine the SN location, as we will see below.  If
we add more detector sites such as Gran Sasso and Sudbury, the results
would be improved.
We note that there is no strong dependence of the results as long as
we select four detector locations which are well separated from each
other.

We note that ANDES is the first deep underground laboratory in the
Southern Hemisphere, which could be constructed in the Agua Negra
tunnels that will link Argentina and Chile under the Andes, the world
longest mountain range.
The potential of a neutrino detector at ANDES location for the observation 
of SN neutrinos as well as of geoneutrinos is discussed in 
Ref.~\cite{Machado:2012ee}.

In Fig.~\ref{fig:two-detectors} we show the solution of 
Eq. (\ref{eq:time-difference}) for the case where the SN occurs 
in the Galactic center, 
given by $\alpha$ = 17h42m27s and $\delta$ = -28$^\circ$55', 
for various different combinations of the four detector sites
mentioned above.
For definiteness, it was assumed that the SN neutrinos
arrived at the Earth on March 20$^{\text{th}}$ of 2000 at 12:00 UTC 
but it is straightforward to change this condition. 

From this plot, we can see that for a given combination of 
two detector sites, the SN location can be constrained, as expected, to a  
closed curve in the sky. 
It is also possible to see that if we have three different  
detector sites, we can restrict the possible SN positions to only two 
locations in the sky. 
For example, the curves for the Kamioka-South Pole 
and Pyh\"asalmi-South Pole combinations intersect in two locations, the 
true location of the SN as well as a {\em fake} solution.
If we have detectors at four different sites, it is possible to 
eliminate the fake solutions and single out the true location 
in the sky, as shown in \cite{Habig-etal-neutrino1998}.
In practice, however, due to the finite resolution of the SN neutrino
arrival time measurement, we can only establish the SN direction with
limited precision.
The accuracy of the determination of $\theta$, the angle between the SN 
direction and the axis connecting two given detectors, 
can be roughly estimated as 
\begin{equation}
\delta (\cos \theta) \sim
\frac{
c\ \delta (\Delta t_{ij}) } {d_{ij}}. 
\end{equation}

Let us try to estimate the precision of the arrival 
time of the SN neutrino signal following the discussion 
given in Ref.~\cite{Beacom:1998fj}.
Let us consider the case where the neutrino event rate $N(t)$ 
at a given detector, which is proportional to 
the SN neutrino flux,  
increases (decreases) before (after) $t=t_0$ exponentially 
as follows, 
\begin{eqnarray}
N(t) =\left\{ \begin{array}{ll}
\propto \exp\left[\displaystyle +\frac{(t-t_0)}{\tau_1}\right] 
& ( t < t_0) \\
&  \\ 
\propto \exp\left[\displaystyle -\frac{(t-t_0)}{\tau_2}\right] 
& ( t >  t_0) \\
\end{array} \right., 
\label{eq:event-rate} 
\end{eqnarray}
where we set $\tau_1 = 30$ ms, 
$\tau_2 = 3$ s following Ref.~\cite{Beacom:1998fj},
and $t_0$ corresponds to the peak of the 
event rate. 
Note that $\tau_1$ ($\tau_2$) characterize the time scale of 
the rising (decaying) part of the time profile of 
the SN neutrino flux or the event rate.
The behavior of the event rate as a function of time 
is shown schematically in Fig. 2 of Ref.~\cite{Beacom:1998fj}.

Under this assumption, very roughly speaking, the accuracy of the
determination of the arrival time of the SN neutrino signal at a given
detector, $\delta t_{\text{arrival}}$, can be estimated
as~\cite{Beacom:1998fj},
\begin{equation}
\delta t_{\text{arrival}} 
\sim \frac{\tau_1 \tau_2}{\sqrt{N}}
\sim \frac{\tau_1}{\sqrt{N_1}},
\label{eq:delta-t-1} 
\end{equation}
where $N_1$ is the number of events in the rising part of the SN
neutrino pulse, given as $N_1 \sim N(\tau_1/\tau_2)$, and $N$ is the
total number of events.  We note that when the event rate is
characterized by Eq. (\ref{eq:event-rate}) with $\tau_1 \ll
\tau_2$, typically the fraction of events relevant for the
determination of $\delta t_{\text{arrival}}$ is only $\sim$ a few \%.

As our reference SN model, we consider the same one 
considered in Ref.~\cite{Machado:2012ee}.  
We assume that the total energy released by neutrinos is 
3$\times 10^{53}$ erg, equally divided by 6 species of neutrinos
and anti-neutrinos. 
We further assume that the SN neutrino spectra are given by 
the parameterization obtained by the Garching 
group~\cite{Keil:2002in,Keil:2003sw,Buras:2002wt},
\begin{eqnarray}
 F^0_{\nu_\alpha}(E) &=& \frac{1}{4 \pi D^2}\ 
 \frac{\Phi_{\nu_\alpha}}{\vev{E_{\nu_\alpha}}}\,
 \frac{\beta_{\alpha}^{\beta_{\alpha}}}{\Gamma(\beta_{\alpha})}  
 \left[\frac{E}{\vev{E_{\nu_\alpha}}}\right]^{\beta_{\alpha}-1}
 \nonumber \\
&\times & \exp\left[-\beta_{\alpha}\frac{E}{\vev{E_{\nu_\alpha}}}\right], \,
\label{eq:flux-Garching}
\end{eqnarray}
where $D$ is the distance to the SN, $\Phi_{\nu_\alpha}$ is the total
number of $\nu_\alpha$ emitted, $\vev{E_{\nu_\alpha}}$ is the average
energy of $\nu_\alpha$ and $\beta_{\alpha}$ is a parameter which
describes the deviation from a thermal spectrum (pinching effect) that can
be taken to be $\sim 2-4$, $\Gamma(\beta_\alpha)$ is the gamma
function. 
As in Ref.~\cite{Machado:2012ee}, 
we set $\beta_\alpha$ = 4 for all flavors, 
$\langle E_{\nu_e} \rangle$ = 12 MeV, 
$\langle E_{\bar{\nu}_e} \rangle$ = 15 MeV, 
$\langle E_{\nu_x}\rangle$ = 18 MeV. 
Here $\nu_x$ implies any non-electron neutrino. 

Due to oscillations, the $\bar{\nu}_e$ SN neutrino spectrum,  
for example, to be observed at the Earth 
gets modified as ~\cite{Dighe:1999bi}, 
\begin{equation}
 F^{\text{obs}}_{\bar{\nu}_e}(E) = 
\bar{p} F^{0}_{\bar{\nu}_e}(E) 
+ (1-\bar{p}) F^{0}_{\nu_x}(E), 
\end{equation}
where $\bar{p}$ is the survival probability of 
${\bar{\nu}_e}$. 
For definiteness and simplicity, 
we consider the neutrino mass hierarchy to be 
normal and ignore any possible effects which could come 
from shock waves (see e.g., \cite{Fogli:2006xy})
and/or non-linear collective effects (see e.g., \cite{Duan:2005cp}).
In this case, to a good approximation~\cite{Dighe:1999bi}, 
we can set $\bar{p} = \cos ^2 \theta_{21} = 0.69$ 
as in ~\cite{Machado:2012ee}. 
We then compute the number of events, $N$, $N_1$, 
and estimate the expected 
uncertainty on the arrival time, $\delta t_{\text{arrival}}$. 
In this work, we consider five detectors, 
2 existing ones, Super-Kamiokande and IceCube, 
one in construction, SNO+, and three others 
which have been proposed, Hyper-Kamiokande~\cite{HK}, LENA~\cite{Wurm:2011zn}
and ANDES~\cite{Machado:2012ee}. 
For the Hyper-Kamiokande detector, for SN neutrino observations, 
we consider the total inner volume of 740 kt. 
For IceCube, we take the numbers estimated by 
the IceCube collaboration~\cite{Abbasi:2011ss}
for a progenitor of 20 M$_\odot$. 
The number of useful neutrino induced 
Cherenkov photons to be recorded by the entire IceCube detector
is $\sim 10^6$. 
For the Water Cherenkov detectors, Super-Kamiokande and 
Hyper-Kamiokande, 
we consider the inverse beta decay $\bar{\nu}_e + p \to n +e^+$, 
and the elastic scattering $\nu_\alpha + e^- \to \nu_\alpha + e^-$, 
whereas for the liquid scintillators,  
SNO+, LENA and ANDES, we consider the inverse beta decay 
and the proton neutrino elastic scattering, 
$\nu_\alpha + p \to \nu_\alpha + p$. 

We show our results in Table \ref{talbe:table-1}
where for a given detector and fiducial mass, 
the total number of events, 
$N$, the number of events in the rising part of the SN signal, 
$N_1$, and the arrival time uncertainty, $\delta t_{\text{arrival}}$,
are shown. 
Our results for IceCube can be compared with 
the ones obtained in ~\cite{Halzen:2009sm}.

According to Ref.~\cite{Halzen:2009sm}, based on Monte Carlo studies,
IceCube can reconstruct the signal onset of the SN neutrinos with a 
resolution of $\delta t_{\text{arrival}} = 1.7$ ms at 1 $\sigma$ C.L., 
for the SN signal consistent with $\tau_1 = 50$ ms.
This value is about a factor 6 worse than what we obtained 
in Table I for IceCube. 
We observe that the difference can be partially 
explained by the fact that Ref.~\cite{Halzen:2009sm} 
considered a larger $\tau_1$ than us and 
also the number of $N_1$ in the first 30 ms 
in Ref.~\cite{Halzen:2009sm} is smaller ($\sim 6 \times 10^3$)
than what we considered here. 
If we simply consider $\tau_1$ = 50 ms and 
$N_1 \sim 6000$, we would obtain 
$\delta t_{\text{arrival}} \sim 0.6$ ms which is 
still smaller than that is obtained in ~\cite{Halzen:2009sm}. 
So it is probably safe to assume that the values obtained 
in Table I could vary within a factor of 2 or so, 
depending on how one estimates.

For the case where the edge is really sharp or if the decaying time 
of the SN signal is considered to be zero, 
roughly corresponding to the case of a failed SN with the 
formation of a black holes (BH), the uncertainty
  on the arrival time is given by the inverse of the event rate before
  the cut off of the SN signal~\cite{Beacom:1998fj},
\begin{equation}
\delta t_{\text{arrival}}^{\text{BH}} \sim \frac{\tau}{N},
\label{eq:delta-t-2} 
\end{equation}
where $\tau$ is the duration of the signal and $N$
is the total number of observed events
to be obtained before the abrupt termination of the
neutrino flux.
According to ~\cite{Sumiyoshi:2006id}
the duration of SN signal before the BH formation 
is $\sim O(1)$ s. 
In this case, for all the detectors considered in 
Table I, except for SNO+, $\delta t_{\text{arrival}}$ 
is less than 1 ms, and 
even for the smallest detector, 
SNO+, we expect that 
$\delta t_{\text{arrival}}^{\text{BH}} 
\sim \tau/N \sim 1/400 = 2.5$ ms. 
We note, however, that due to the 
uncertainty associated with the formation of the black hole, 
which is about 0.5 ms~\cite{Beacom:2000qy}, 
$\delta t_{\text{arrival}}^{\text{BH}}$ can not 
be smaller than 0.5 ms. 

\begin{table}[t!]
\begin{tabular}{lcccc}
\hline
Detector &  \ Fid. Mass (kt)&  $N$\  & \ $N_1$ \ 
& \ $\delta t_{\text{arrival}}$\ (\text{ms}) \\
\hline\hline
Super-K\  & 32 &  8.0 $\times 10^3$ & 80  & 3.4 \\ 
Hyper-K & 740& 1.9$\times 10^5$  &  1.9 $\times 10^3$ & 0.7 \\
SNO+    & 0.8& 400  & 4& 15 \\
LENA    & 44 & 1.8$\times10^4 $  & 1.8$\times10^2$ & 2.7 \\
ANDES    & 3 & 1.2$\times10^3$   & 12 &  8.7 \\
IceCube    & $\sim 10^{3}$& $\sim 10^{6}$  & $\sim 10^{4}$ & 0.3 \\
\hline
\end{tabular}
\caption{\label{tab1} 
Estimated number of events for a SN at 10 kpc from the Earth,
as well as the expected precision on the arrival time 
of the SN signal for the existing detector, 
IceCube, Super-Kamiokande (denoted as Super-K), 
as well as the proposed neutrino 
detectors, SNO+ (in construction), 
LENA, Hyper-Kamiokande (denoted as Hyper-K), 
and ANDES.}
\label{talbe:table-1} 
\end{table} 

\section{Combined Analysis}
\label{sec:results}

In this section, we discuss the results of our combined analysis by
considering observations of SN neutrinos at three and four different
detector sites on the Earth.

We define our $\chi^2$ function as follows,
\begin{equation}
\chi^2  = \sum_{i,j}
\left[ \frac{ \Delta t_{ij}^{\text{obs}} (\alpha_0,\delta_0) -  
\Delta t_{ij}^{\text{theo}}(\alpha,\delta)  }
{\sigma_{\Delta t}}\right]^2,
\end{equation}
where $\Delta t_{ij}^{\text{obs}}(\alpha_0,\delta_0)$ 
is the arrival time difference 
of SN neutrinos to be observed (expected) for the input (true) 
SN location in the sky $(\alpha_0,\delta_0)$ for the 
combination of $i$-th and $j$-th detector sites on the Earth
whereas $\Delta t_{ij}^{\text{theo}}(\alpha,\delta)$ is 
the theoretically expected one for a given SN location $(\alpha,\delta)$. 
$\sigma_{\Delta t}$ is the assumed time resolution. 
Note that by construction, the best fit values $(\alpha,\delta)$ 
obtained by our $\chi^2$ analysis are the solution of 
the Eq. (\ref{eq:time-difference}) for an input value of  
$\Delta t_{ij}^{\text{obs}}$. 

\begin{figure*}[!t]
\begin{center}
\hglue -0.3cm
\includegraphics[width=0.90\textwidth]{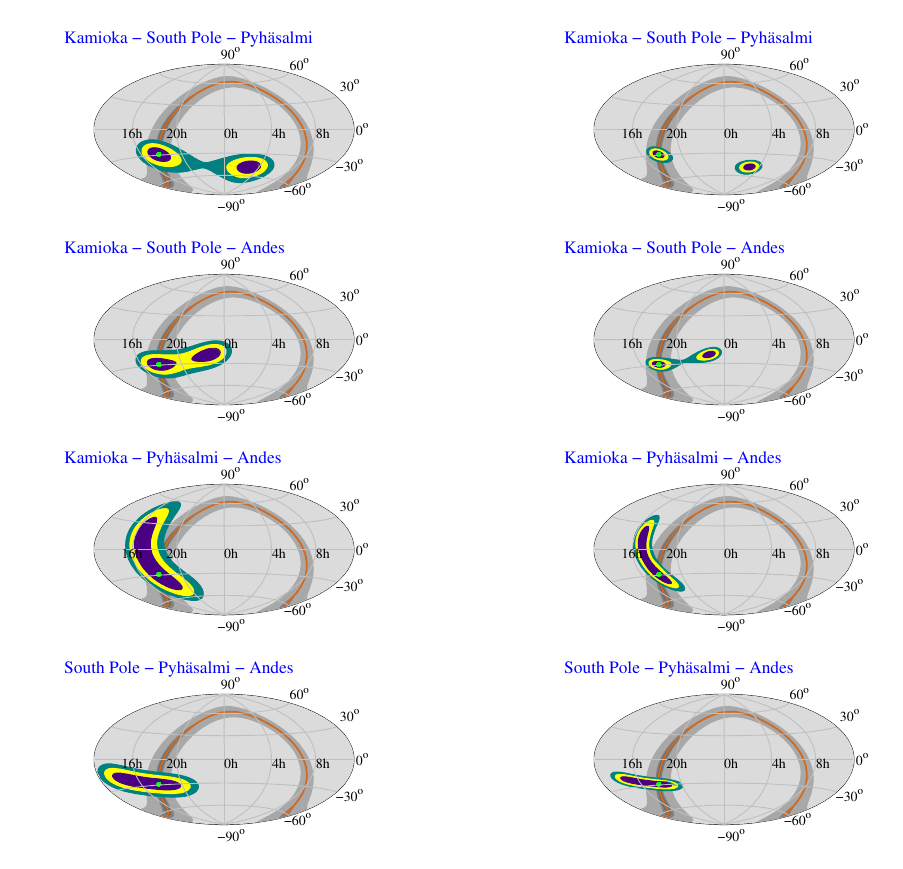} 
\end{center}                          
\vglue -1.0cm
\caption{ Cases where three detectors are considered to determine the
  position of the Galactic SN: Kamioka-South Pole-Pyh\"asalmi (first
  row), Kamioka-South Pole-ANDES (second row),
  Kamioka-Pyh\"asalmi-ANDES (third row), and South
  Pole-Pyh\"asalmi-ANDES (fourth row).  The colors purple, yellow and
  green indicate, respectively, the regions allowed at 1, 2 and 3
  $\sigma$ C.L..  Here the uncertainty in the time difference
  measurement between two detectors is assumed to be $\pm$ 4 ms (left
  panels) $\pm$ 2 ms (right panels).  }
\label{fig:three-detectors}
\end{figure*}

In Fig.~\ref{fig:three-detectors} we show for the
same input SN location at the Galactic center used in
Fig.~\ref{fig:two-detectors} what would be the angular resolution 
for ($\alpha$, $\delta$) that would
result from a combination of arrival time differences registered by
three different detectors.  
For definiteness and simplicity,
we assume, for the
combination of two detectors, that the arrival time difference
resolution can be $\pm $ 4 ms (left panels) and 
$\pm$ 2 ms (right panels).


\begin{figure*}[!t]
\begin{center}
\hglue -0.3cm
\includegraphics[width=1.0\textwidth]{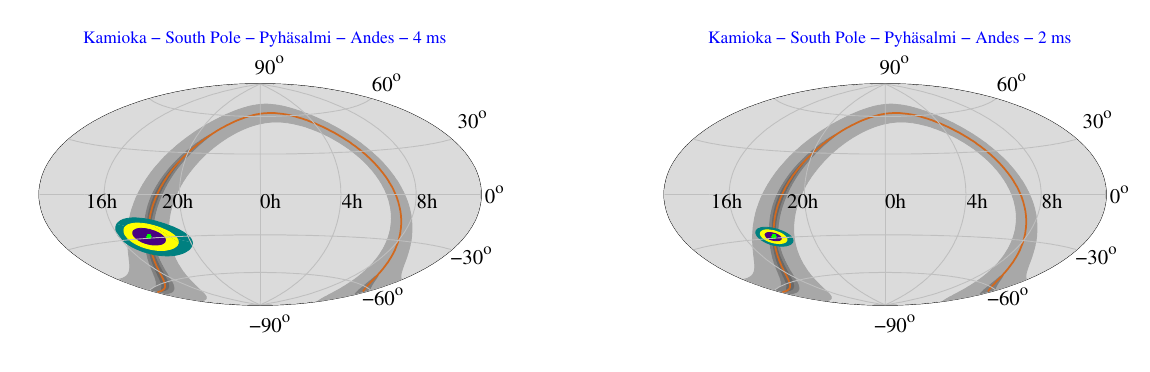}
\end{center}                          
\vglue -1.1cm
\caption{
Allowed regions at 1, 2 and 3 $\sigma$ C.L. 
compatible with the combinations of the arrival time differences assuming 
four detector sites: Kamioka, the South Pole, Pyh\"asalmi and ANDES.
We assumed the SN to be at the Galactic center and that the uncertainty 
in the time difference measurement between two detectors to be $\pm$ 4 ms
(left panels) and $\pm$ 2ms (right panels).}
\label{fig:four-detectors-A}
\end{figure*}

\begin{figure*}[!t]
\begin{center}
\hglue -0.3cm
\includegraphics[width=1.0\textwidth]{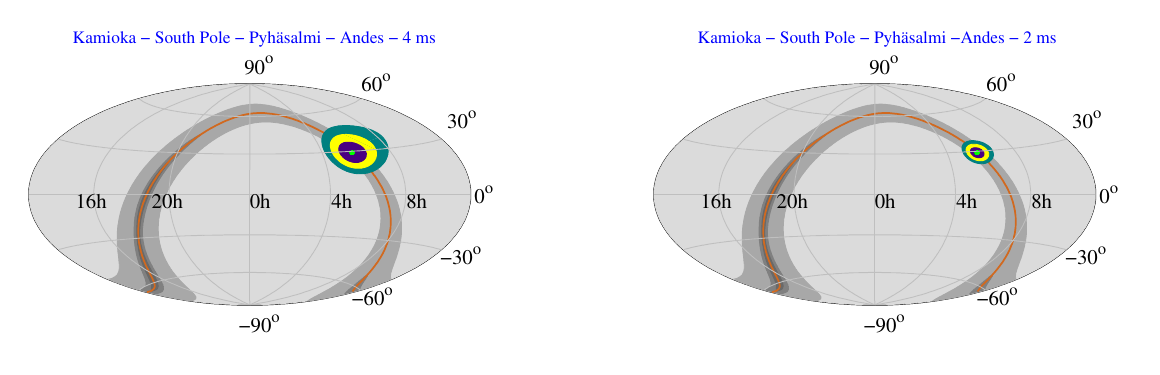}
\end{center}                          
\vglue -1.1cm
\caption{
Same as Fig.~\ref{fig:four-detectors-A} but for 
a SN explosion that occurred at a location opposite to the Galactic
center.} 
\label{fig:four-detectors-B}
\end{figure*}

As expected, for all the combinations we considered, we obtained 
two solutions at different locations in the sky, the true solution 
and the fake one.
We note the true and fake allowed regions are 
connected at 1 $\sigma$ C.L. for
the cases shown in the lower four panels in 
Fig.~\ref{fig:three-detectors}.
Though we have two solutions, in practice, the one that lay in the region 
of the Galactic disc has greater probability of being the true one.
The fake solution can be eliminated 
by considering a fourth detector location as we 
can see below. 

In Fig.~\ref{fig:four-detectors-A} we show the case where 
four different detector locations are considered for the 
same SN input location in the Galactic center for the time resolution 
of $\pm$ 4 ms (left panels) and $\pm$ 2 ms (right panels). 
With four detectors at different sites, it is 
possible to single out the true location of the SN.
In Fig.~\ref{fig:four-detectors-B} we show  similar plots
for the case of different SN input location, opposite to the 
Galactic center, 
$\alpha$ = 5h42m27s and $\delta$ = 28$^\circ$55'.  
From these plots we can conclude that the 
expected precision is $\Delta (\alpha) \sim  15 (8)^\circ$ 
and $\Delta (\delta)  \simeq  10 (5)^\circ$ for 
the time resolution of $\pm$ 4 ms (2 ms).

Let us make a brief summary of our results on how the pointing
accuracy depends on the number of detectors used in the analysis.  If
we consider only two detectors, in general, the region compatible with
the location of the SN in the sky is quite large, as we can easily
guess from Fig.~\ref{fig:two-detectors} although this plot corresponds
to the case of no arrival time uncertainty or $\delta
t_{\text{arrival}}=0$.  From two to three detectors, the reduction of
the region compatible with the SN location is quite sizable, we can
see this by comparing Figs.~\ref{fig:two-detectors} and
\ref{fig:three-detectors}.  From three to four detectors, roughly
speaking, the region is again reduced by about one half, as we can see
by comparing Figs.~\ref{fig:three-detectors},
\ref{fig:four-detectors-A}, and \ref{fig:four-detectors-B}.

We note, however, that even if we consider four detectors, the 
allowed angular ranges of the SN location in the sky are much larger than 
the typical field of view of an optical telescope. 
Nevertheless, this is better than no information at all and 
especially useful in the case where the SN can not be 
located by optical observation, because of dust,
or, in the case of a failed SN, because it is accompanied by a BH 
formation.

\section{Conclusions}
\label{sec:conclusions}

The era of high-statistics neutrino detectors has started.  IceCube is
already operating in the South Pole and in a decade or so we expect to
also have Hyper-Kamiokande in Japan as well as one very large neutrino
detector in the future European underground laboratory in 
Pyh\"asalmi. There is also a possibility to construct 
a new neutrino detector in the Southern Hemisphere at ANDES.
This makes the determination of the angular position of a nearby SN by
comparing the arrival time of the first SN neutrinos at these 
different detector locations on the Earth an interesting possibility.  

The time resolution of the triangulation technique will be dominated
by the smallest detector, since the precision of the reconstruction of
the neutrino signal onset depends on the number of
neutrinos registered by the detector (see Eqs. (\ref{eq:delta-t-1}) 
and (\ref{eq:delta-t-2})).
We have demonstrated that, in general, one needs to combine the timing 
of four different detector locations in order to uniquely localize the 
SN using this method.

Assuming a rather optimistic, but not impossible, uncertainty on the
arrival time difference between two detectors to be $\sim \pm$ (2-4)
ms, we have estimated the angular resolution of the determination of
the location of a SN that could occur in the Galactic center given by
four detectors located at the South Pole (IceCube), Kamioka
(Super-Kamiokande or Hyper-Kamiokande), 
ANDES and Pyh\"asalmi (LENA, MEMPHYS and GLACIER).  
We established that in this case the angular position
can be known within $\sim$ 5 (10)$^\circ$ in declination and $\sim 8
(15)^\circ$ in right ascension for the time resolution of 2 (4) ms.

\begin{acknowledgments} 
  \vspace{-0.3cm} This work is supported by Funda\c{c}\~ao de Amparo
  \`a Pesquisa do Estado de S\~ao Paulo (FAPESP), Funda\c{c}\~ao de
  Amparo \`a Pesquisa do Estado do Rio de Janeiro (FAPERJ) and by
  Conselho Nacional de Ci\^encia e Tecnologia (CNPq).  R.Z.F. also
  thanks Institut de Physique Th\'eorique of CEA-Saclay for the
  hospitality during the time this work was developed and acknowledges 
  partial support from the European Union  FP7 ITN 
  INVISIBLES (PITN-GA-2011-289442).
\end{acknowledgments}

\appendix 

\section{Description of the Equatorial Coordinate System}
\label{sec:coord}

The SN location can be given in the so-called Equatorial Coordinate
System by two angular coordinates, $\alpha$ known as right ascension
and $\delta$ known as declination. Right ascension measures the
angular distance eastward along the celestial equator from the vernal
equinox, its is analogous to terrestrial longitude. Usually right
ascension is not given in degrees but rather in sidereal hours, minutes
and seconds.  The vernal point is defined by where the celestial
equator and the ecliptic intersect at 00$^{\rm h}$00$^{\rm m}$00$^{\rm
  s}$ and longitude 0$^\circ$.  By definition the north celestial pole
corresponds to $\delta=+90^\circ$, so it is analogous to the
terrestrial latitude.  

This defines the unit vector ${\bf n_0}$, which
points in the direction of propagation of the neutrinos arriving at
the Earth coming from the SN as
\begin{equation}
{\bf n_0} = (n_{0x}, n_{0y}, n_{0z}),
\label{Eq:n0}
\end{equation}
where 
\begin{eqnarray}
n_{0x} & = &  - \cos \alpha \sin \delta \nonumber \\
n_{0y} & = &  - \sin \alpha \sin \delta \nonumber \\
n_{0y} & = &  - \cos \delta. 
\label{Eq:SN-direction}
\end{eqnarray}

Let us assume that a detector positioned at the $i$-th site
on the Earth is localized, 
at a certain time $t$, by the following vector 
\begin{equation}
{\bf x}_{i} = (x_i, y_i, z_i), 
\end{equation}
with coordinates 
\begin{eqnarray}
x_i(t) & = &  R_\oplus \cos \phi_i(t)\sin \theta_i \nonumber \\
y_i(t) & = &  R_\oplus \sin \phi_i(t) \sin\theta_i \nonumber \\
z_i(t) & = &  R_\oplus \cos \theta_i,
\end{eqnarray}
where  $R_\oplus$ is the radius of the Earth and $\theta_i$ is the 
latitude corresponding to the position of the detector. 
The angle $\phi_i(t)$ depends on time and can be given by 
\begin{eqnarray}
\phi_i(t) & = &  \phi_i(0) + \omega \, t -\Omega \, T -\pi,
\end{eqnarray}
where $\phi_i(0)$ is the longitude  corresponding to the initial 
position of the detector, $\omega$ is the angular velocity 
of the daily rotation of the Earth and $\Omega$ is the angular velocity 
corresponding to the annual rotation of the Earth around the Sun. 
The time $t$ refers to the moment of the day the SN explosion occurred
($0 \le t \le 24$ h),  given in terms of the Coordinated Universal Time (UTC), 
whereas $T$, assumed to be common for all detectors,  is the time elapsed 
after the vernal point when the detector received the SN neutrinos. 

So, if we have two detectors, say, one at site 1 and the other at site 2, 
we can, explicitly, write the observed arrival time difference as 
\begin{eqnarray}
\Delta t_{12}  &= &
(R_\oplus/c) [
(\cos \phi_1(t)\sin \theta_1 -\cos \phi_2(t)\sin \theta_2)n_{0x} \nonumber \\
&+& (\sin \phi_1(t)\sin \theta_1 -\sin \phi_2(t)\sin \theta_2)n_{0x} \nonumber \\
&+& (\cos\theta_1 - \cos \theta_2)n_{0z} ],
\label{Eq:Delta-t2}
\end{eqnarray}
which constrains the possible values of $\alpha$ and $\delta$.

\end{document}